\documentclass[12pt]{article}

\usepackage[T1]{fontenc}
\usepackage[utf8]{inputenc} 
\usepackage{authblk}
\usepackage{hyperref}
\usepackage{graphicx}
\usepackage{amsmath,amssymb}
\usepackage[x11names]{xcolor}
\usepackage{wasysym}

\newcommand{\seff}[1]{\sin^2\vartheta_{\rm eff}^{#1}}

\def\phAV{\phantom{ AVALUE =}}
\newcommand{\sss}[1]{\scriptscriptstyle{#1}}
\def\als{\alpha_s}
\def\GeV{\unskip\,\mathrm{GeV}}
\def\MeV{\unskip\,\mathrm{MeV}}

\begin{document}

\title{Computer package {\tt DIZET} v.~6.45}

\author[a]{A.~Arbuzov}
\author[b]{J.~Gluza}
\author[a]{L.~Kalinovskaya}
\author[c]{S.~Riemann}
\author[c]{T.~Riemann}
\author[a,d]{V.~Yermolchyk}

\affil[a]{\small Dubna, 141980 Russia}
\affil[b]{\small Chorz\'ow, 41-500 Poland}
\affil[c]{\small {\color{black}K\"onigs Wusterhausen, D-15711}
Germany }
\affil[d]{\small Minsk, 220006 Belarus}

\maketitle
\begin{abstract}
The new version of the {\tt DIZET} electroweak library is described.
Changes and additional code features concerning the 
previous version are explained. The software allows one to make 
state-of-the-art theoretical predictions for pseudo-observable quantities, 
including higher-order radiative corrections.
The current version of the {\tt DIZET} library v.~6.45 incorporates 
advanced recent results of theoretical calculations. 
Numerical comparisons with the results of the previous version are performed.
Estimates of theoretical uncertainties are discussed.
\end{abstract}
 
\section{Introduction}

The computer package {\tt DIZET} was created as
electroweak and QCD library of 
the {\tt ZFITTER} program~\cite{Bardin:1999yd,Arbuzov:2005ma} which was one of the
main tools for the high-precision verification of the Standard Model at LEP~\cite{ALEPH:2005ab}.
{\tt DIZET} can also be linked as 
a library by other projects, e.g., it is used by the {\tt HECTOR} program~\cite{Arbuzov:1995id}
and by the {\tt KKMC} Monte Carlo event generator~\cite{Arbuzov:2020coe}.

{\tt DIZET} can be used for fitting EWPOs, for instance $\sin^2\vartheta_{eff}$ at LHC as discussed at the LHC EW precision workshop \cite{dizetws}. 
In each new version of {\tt DIZET}, the compatibility with all previous versions has been preserved.
Thus, the numerics of a previous version can be fully reproduced,
except for changes caused by the correction of bugs. The latter is documented
in the header of the code.
We remind, that between {\tt DIZET} versions 6.21~\cite{Bardin:1999yd} and 
6.42~\cite{Arbuzov:2005ma}, there are changes affecting the $W$ boson width and the running of 
the electromagnetic coupling $\alpha_{QED}$.
In addition, the treatment of so-called box-like diagram contributions 
(controlled by the IBOXF flag of the {\tt ROKANC} subroutine in {\tt ZFITTER}) 
and of the $b$ quark production channel (IBFLA flag in the same subroutine) 
have been changed.
Links to publications and the public versions of {\tt DIZET} can be found at the {\tt ZFITTER} project webpage 
\href{http://sanc.jinr.ru/users/zfitter/}{http://sanc.jinr.ru/users/zfitter/}.

The last documented version of {\tt DIZET} is
6.42~\cite{Arbuzov:2005ma}.
In the present paper, we describe the actual {\tt DIZET} version 6.45~\cite{dizetv645}. 
We present the transition from  {\tt DIZET} v.~6.42 to {\tt DIZET} v.~6.45 and show the numerical impact of the newly introduced modifications, controlled by the corresponding options and flags on pseudo-observables (EWPOs). 
The contributions added in  {\tt DIZET} v.~6.45 are connected with the completion of the 2-loop EW radiative corrections given in \cite{Dubovyk:2016aqv,Dubovyk:2019szj}
and which complement earlier works on radiative corrections, namely: the complete  fermionic  two-loop corrections to the $W$ boson mass \cite{Freitas:2002ja}; the  leading  ${\cal  O}(\alpha  \als)$
\cite{
Djouadi:1993ss}
and next-to-leading ${\cal O}(\alpha\als^2)$
\cite{Avdeev:1994db,Chetyrkin:1995ix,Chetyrkin:1995js}          
QCD corrections,  as well as  leading three-loop corrections  in an expansion in  $m_t^2$ of order   ${\cal  O}(\alpha^3)$   and   ${\cal    O}(\alpha^2  \als)$
\cite{Faisst:2003px}. 

These modifications are relevant for future precision HL-LHC studies and the LHC electroweak Working Group activities.
They are also needed as a first step towards high precision predictions of the Standard Model electroweak effects 
at future high energy colliders. In the context of the future circular electron-positron collider (FCC-ee)~\cite{Abada:2019zxq}, 
the anticipated experimental accuracy on EWPOs has to be matched with theory predictions of at least the same level of accuracy to achieve maximum usage of experimental data. 
For the present situation concerning EWPOs determination and their future estimate, see Tab.~\ref{tab:th} and references \cite{Blondel:2018mad,Freitas:2019bre} for more details. In Tab.~\ref{tab:th}, we put experimental predictions for EWPOs at FCC-ee as the most stringent among future experimental setups, particularly in the Z-resonance region.  
Other  widely considered future $e^+e^-$ collider projects are CEPC~\cite{CEPCStudyGroup:2018ghi}, ILC~\cite{Baer:2013cma,Bambade:2019fyw}, and CLIC~\cite{Linssen:2012hp,Charles:2018vfv}. 
Through their high integrated luminosities of several ab$^{-1}$  (practically for all relevant Z-resonance, $HZ$, $WW$, and $t\bar{t}$ modes \cite{Abada:2019zxq,CEPCStudyGroup:2018ghi}) these machines will be sensitive to very small deviations between the measured value and the SM expectation 
for a given observable.
To account correctly for such slight deviations, dedicated programs like here discussed  
{\tt DIZET} will be highly needed.
Table~\ref{tab:th} shows the  comparison between the estimated FCC-ee experimental precision, the current theoretical uncertainty, 
and the so-called projected one for
representative EWPOs, see Chapter~B in~\cite{Blondel:2018mad} and~\cite{Freitas:2019bre}. 
By the projected theoretical uncertainty we mean an estimate of the future theoretical uncertainty when the leading 3-loop $\mathcal{O}(\alpha^3, \alpha^2\alpha_s, \alpha\alpha_s^2)$ 
corrections will become available.

\begin{table}[ht]
    \centering
    \begin{tabular}{|l|ccc|}
    \hline
Quantity & FCC-ee & Current theory
& Projected theory \\
  &  & uncertainty &
uncertainty \\  \hline 
$m_\mathrm{W}$ (MeV) & $0.5-1$ & 4
& 1  \\
    $\sin^2 \vartheta^\ell_{\rm{eff}}$ ($10^{-5}$) & 0.6 & 4.5
    & 1.5  \\
    $\Gamma_\mathrm{Z}$ (MeV) & 0.1 & 0.4 
    & 0.15  \\
    $R_b$ ($10^{-5}$) & 6 & 11 
    & 5  \\
    $R_\ell$ ($10^{-3}$) & 1 & 6 
    & 1.5  \\
   \hline
    \end{tabular}
\caption{Estimated precision for the direct determination of representative EWPOs at FCC-ee (column 2), current theory uncertainties 
for the SM prediction of these quantities (column 3), and the projected theoretical uncertainty (column 4).}
    \label{tab:th}
\end{table}

Indeed, as {\tt DIZET} includes Standard Model higher order radiative corrections, it can be used for comparisons with experimental results in search for models which go beyond the Standard Model. We discuss here the impact of newly implemented corrections in {\tt DIZET}
on EWPOs and form-factors at the $e^+e^-$ resonance.

\section{Release {\tt DIZET} from v.~6.42 to v.~6.45}

The Fortran code {\tt DIZET} is a library for the calculation of
electroweak radiative corrections and it is part of the {\tt ZFITTER} distribution package.
It can also be used in a stand-alone mode.

On default, {\tt DIZET} performs the following calculations:
\begin{itemize}
\item by call of subroutine {\tt ROKANC}: four weak neutral-current (NC) 
form factors, running electromagnetic 
 and strong couplings needed for the calculation of effective NC Born 
cross sections for the production of massless fermions (however, the mass
of the top quark appearing in the virtual state of loop diagrams
for the process $e^+ e^- \to f \bar{f}$ is not ignored); 
\item
by call of subroutine {\tt RHOCC}: the corresponding form factors and running 
strong coupling for the calculation of effective CC Born cross sections; 
\item
by call of subroutine {\tt ZU\_APV}: $Q_W(Z,A)$ -- the weak charge used for
the description of parity violation in heavy atoms.
\end{itemize}
If needed, the form factors of cross sections may be made to contain
the contributions from  
$WW$ and $ZZ$ box diagrams thus ensuring  the correct kinematic
behaviour over a larger energy range compared to the Z pole.

Between the {\tt DIZET} versions 6.42 and 6.45 there are changes affecting the running 
of the QED coupling $\alpha(s)$ and the QCD corrections of
order  
$\alpha\alpha_S$ to the $Z$ boson partial widths.
Also, starting from v.6.44, {\tt DIZET} uses the complete $\alpha_s^4$ QCD corrections to hadronic $Z$-decays \cite{Baikov:2012er} by default. In order to reproduce the old behavior, the IBAIKOV flag 
must be set to 2014 in the code of {\tt DIZET} v.~6.45.
The largest contribution of the electroweak (EW) corrections 
comes from the $s$ channel QED running of $\alpha(s)$, and 
the main load in it is due to the hadronic component
$\Delta\alpha^{(5)}_{had}(M_{\sss Z})$~\cite{Jegerlehner:2003ip}.

\subsection{New options in {\tt DIZET} v.~6.45}
    
In this section, we give the descriptions 
of flags and 
added options implemented in {\tt DIZET} v.~6.45.

\vspace*{0.4cm}

$\bullet$ flag {\bf IAMT4:}

\begin{description}
\item{\underline{\bf IAMT4:}
two-loop $\alpha^2$ bosonic and/or fermionic radiative corrections: }
\begin{description}
\item[IVALUE =~~I]
\item[\phAV 6] --- fermionic two-loop corrections to  $\seff{l}$~\cite{Awramik:2004qv};
\item[\phAV 7]  ---   the complete two-loop corrections to $\seff{b}$ and $\seff{l}$ according to 
Refs.~\cite{Dubovyk:2016aqv,Awramik:2006uz}. 
\item[\phAV 8]  ---  the complete electroweak two-loop radiative corrections to all the 
relevant electroweak precision pseudo-observables related to the Z-boson,
according to Ref.~\cite{Dubovyk:2019szj}.
\end{description}

\end{description}

The complete set of EWPOs related to the $Z$-boson for
IAMT4 = 8  includes: 
the leptonic and bottom-quark effective weak mixing angles
$\seff{\ell}$, $\seff{b}$, 
the $Z$-boson partial
decay widths $\Gamma_f$, where $f$ indicates any charged lepton, neutrino and quark flavor (except for the top quark), the total $Z$ decay width $\Gamma_Z$, the 
branching ratios $R_\ell$, $R_c$, $R_b$, and the hadronic cross section $\sigma_{\rm had}^0$.

\vspace*{0.4cm}

$\bullet$ flag {\bf IHVP:}

\begin{description}
\item{\underline{\bf IHVP} --- choice of hadronic vacuum polarization $\Delta \alpha^{(5)}_{had}(M_{\sss Z})$
using public versions of the {\tt AlphaQED} code by F.~Jegerlehner:}
\begin{description}
\item[IVALUE =~~I]
\item[\phAV 1] --- realization of the fit given in \cite{Eidelman:1995ny},
\item[\phAV 4] --- realization 
of the fit by  \cite{Jegerlehner:2015stw},
\item[\phAV 5] --- realization of the fit by \cite{Jegerlehner:2017zsb}.
\end{description}
\end{description}

Details on hadronic vacuum polarization effects 
can be found in~\cite{Jegerlehner:2017zsb}.

\begin{table}[ht]
\centering
\begin{tabular}{|c|c|c|c|}
  \hline
IHVP                        & 1         &   4       &    5       \\
	\hline
version	                     & FJ-1995   & FJ-2016   &   FJ-2017  \\
	\hline
$\Delta \alpha^{(5)}_{had}(M_{\sss Z}) $ & 2.8039e-2  & 2.7586e-2 &   2.7576e-2\\
	\hline
\end{tabular}
\caption{Results of the fit for hadronic vacuum polarization 
$\Delta \alpha^{(5)}_{had}(M_{\sss Z})$
for different versions 
(1995-\cite{Eidelman:1995ny}, 
2016-\cite{Jegerlehner:2015stw}, 
2017-\cite{Jegerlehner:2017zsb})}
\end{table}

Note that the  {\tt AlphaQED} (2017) code provides an estimation of statistical and systematic errors.
To estimate the resulting uncertainties of a {\tt DIZET} output one has to run the code in a cycle with
variation of the input parameters such as the top quark and Higgs boson masses within their error bars,
see, e.g., Ref.~\cite{Bardin:1999gt}. In addition one has to estimate the missing contributions of not
yet computed higher order corrections.

\section{Numerical results}

All numbers presented below are obtained with the following set of Input Parameters (IPS)
and their variations within experimental errors taken from PDG Summary Tables, ~\cite{ParticleDataGroup:2020ssz}:
$\alpha^{-1}(0) = 137.035999084$, 
$\alpha_s(M_Z) = 0.1179$, 
$M_Z = 91.1876 \; \mathrm{GeV}$, 
$M_H = 125.25 \; \mathrm{GeV}$,
$m_t = 172.76 \; \mathrm{GeV}$. 
The masses of the five light quarks are chosen in the usual way to
reproduce the hadronic contribution to the photon vacuum polarization (relevant only for $IHVP=2$)
$\Delta \alpha^{(5)}_{had}(M_{\sss Z})$.
Here numerical calculations were carried out at the fixed value IBAIKOV=2014.

The numerical results presented here are slightly different from those of our 
report~\cite{dizetws} due to the change of input parameters.
The present values of pseudo-observables (EW boson widths and the weak mixing angle) are~\cite{ParticleDataGroup:2020ssz}:
$\Gamma_Z   = 2495.200 \pm 2.300$~MeV,
$G_Z(\mu\mu) = 83.99 \pm 0.16$~MeV,
$\Gamma_W =2085 \pm 42$~MeV,
$G_W(l\nu) =  226.4 \pm 1.9$~MeV,
$\sin^2 \vartheta_{eff}\times 10^6= 231480 \pm 160$.

\subsection{Parametric uncertainties}

{\tt DIZET} can calculate pseudo-observables and EW form-factors in a wide range of input parameters: 
$M_{\sss H}$, $M_{\sss Z}$, $m_t$, $\alpha_s$.
Tables~\ref{tablemt}
- \ref{tableas} present the  dependence of pseudo-observables on the experimental uncertainty of input parameters: ($m_t=172.76(0.30) \GeV$, $M_{\sss H}=125.25(0.17) \GeV$, $M_{\sss Z}=91.1876(0.0021) \GeV$, $\alpha_s=0.1179(0.0009)$).
The first type of theoretical uncertainties are due to variation of input parameters within 
experimental errors.
We consider first the 
parametric uncertainties due to variation of the masses $m_t$, $M_{\sss H}$ and $M_{\sss Z}$ 
and in addition the  $\alpha_s$-dependence.

\begin{table}[ht]
\centering
\begin{tabular}{|l|l|l|l|l|}
\hline
$m_t,~\GeV$ & $172.76-0.30$ & 172.76 & $172.76+0.30$ & Diff.\\
\hline
$G_Z(\mu\mu),~\MeV$& 83.982  & 83.985 & 83.987 & 0.005\\
\hline
$\Gamma_Z,~\MeV$ & 2494.746  & 2494.814 & 2494.883 & 0.137\\
\hline
$G_W(l\nu),~\MeV$ & 678.935  & 678.981 & 679.027 & 0.092\\
\hline
$\Gamma_W,~\MeV$ & 2089.825  & 2089.967 & 2090.109 & 0.284\\
\hline
$\sin^2 \vartheta^l_{eff}\times 10^6$ & 231508  & 231500 & 231491 & 17\\
\hline
\end{tabular}
\caption{The effect of the parametric uncertainty in $m_t$
on the magnitudes of pseudo-observables.}
\label{tablemt}
\end{table}
As one can see, the parametric uncertainties for the listed pseudo-observables
are less than the current experimental errors~\cite{ParticleDataGroup:2020ssz}.

\begin{table}[h!]
\centering
\begin{tabular}{|l|l|l|l|l|}
\hline
$M_{\sss H},~\GeV$ & $125.25-0.17$ & 125.25 & $125.25+0.17$ & Diff.\\
\hline
$G_{\sss Z}(\mu\mu),~\MeV$ & 83.985  & 83.985 & 83.985 & 0\\
\hline
$\Gamma_{\sss Z},~\MeV$ & 2494.818  & 2494.814 & 2494.811 & 0.007\\
\hline
$G_{\sss W}(l\nu),~\MeV$ & 678.983  & 678.981 & 678.979 & 0.004\\
\hline
$\Gamma_{\sss W},~\MeV$& 2089.973  & 2089.967 & 2089.961 & 0.012\\
\hline
$\sin^2 \vartheta^l_{eff}\times 10^6$ & 231499  & 231500 & 231500 & 1\\	
\hline
\end{tabular}
    \caption{
    The effect of the parametric uncertainty in $M_{\sss H}$
on the magnitudes of pseudo-observables.}
\label{tablemh}
\end{table}

Tables~\ref{tablemt} and~\ref{tablemh} show that the effect of experimental uncertainty of $m_t$ and $M_{\sss H}$
changes the partial widths by an interval not exceeding their experimental errors.

\begin{table}[ht]
\centering
\begin{tabular}{|l|l|l|l|l|}
\hline
$M_{\sss Z},~\GeV$ & $91.1876-0.0021$ & 91.1876 & $91.1876+0.0021$ & Diff.\\
\hline
$G_{\sss Z}(\mu\mu),~\MeV$ & 83.978 & 83.985 & 83.991 & 0.013\\
\hline
$\Gamma_{\sss Z},~\MeV$ & 2494.602 & 2494.814 & 2495.027 & 0.425\\
\hline
$G_{\sss W}(l\nu),~\MeV$ & 678.914 & 678.981 & 679.048 & 0.287\\
\hline
$\Gamma_{\sss W},~\MeV$ & 2089.761 & 2089.967 & 2090.173 & 0.412\\
\hline
$\sin^2 
\vartheta^l_{eff}\times 10^6$ & 231515  & 231500 & 231485 & 30\\
\hline
\end{tabular}
\caption{The effect of the parametric uncertainty in $M_{\sss Z}$
on the magnitudes of pseudo-observables.}
\label{tablemz}
\end{table}

\begin{table}[!ht]
\centering
\begin{tabular}{|l|l|l|l|l|}
\hline
$\alpha_s$ & $0.1179-0.0009$ & 0.1179 & $0.1179+0.0009$ & Diff.\\
\hline
$G_{\sss Z}(\mu\mu),~\MeV$ & 83.985 & 83.985 & 83.984 & 0.001\\
\hline
$\Gamma_{\sss Z},~\MeV$ & 2494.338 & 2494.814 & 2495.290 & 0.952\\
\hline
$G_{\sss W}(l\nu),~\MeV$ & 678.995 & 678.981 & 678.967 & 0.028\\
\hline
$\Gamma_{\sss W},~\MeV$ & 2089.607 & 2089.967 & 2090.326 & 0.719\\
\hline
$\sin^2 \vartheta^l_{eff}\times 10^6$ & 231497  & 231500 & 231503 & 6\\
\hline
\end{tabular}
\caption{The effect of the parametric uncertainty in $\alpha_s$
on the magnitudes of pseudo-observables.}
\label{tableas}
\end{table}
As seen from Tables \ref{tablemt} $-$ \ref{tablemz}, the largest uncertainty comes to $\Gamma_{\sss W}$ and $\Gamma_{\sss Z}$ due to errors in $M_{\sss Z}$.
These parametric uncertainties remain, however, well below the corresponding experimental errors.

\subsection{Impact of new options}

Numerical results for the comparison of versions
are conveniently presented as
difference in values
for a given observable at different 
sets of flags IHVP and IAMT4.

\subsubsection{Partial $G_{ij}$ and total $\Gamma_{tot}$ decay widths of the Z-boson} 

\begin{table}[!h]
\centering
\begin{tabular}{|l|ccc|c|c|}
\hline         
IHVP,~IAMT4             &      1,8  &        5,6   &      5,8
                                                  & $|\delta_{(1,8)-(5,8)}|$&$|\delta_{(5,6)-(5,8)}|$  \\
\hline
channel                &           &              &            &  &                                   \\
\hline 
 $G_{\nu,\bar\nu},~\MeV$      &   167.202  &   167.202  &    167.202  & 0      & 0      \\
 $G_{e^+,e^-},~\MeV$         &    83.977  &    83.984  &     83.985  & 0.008  & 0.001  \\
 $G_{\mu^+,\mu^-},~\MeV$    &    83.977  &    83.983  &     83.985  & 0.008  & 0.002  \\
 $G_{\tau^+,\tau^-},~\MeV$   &    83.787  &    83.794  &     83.795  & 0.008  & 0.001 \\
 $G_{u,\bar u},~\MeV$       &   299.832  &   299.902  &    299.918  & 0.086  & 0.016  \\
 $G_{d,\bar d},~\MeV$       &   382.783  &   382.846  &    382.861  & 0.078  & 0.015  \\
 $G_{c,\bar c},~\MeV$        &   299.766  &   299.836  &    299.852  & 0.086  & 0.016  \\
 $G_{s,\bar s},~\MeV$       &   382.783  &   382.846  &    382.861  & 0.078  & 0.015  \\
 $G_{b,\bar b},~\MeV$      &   375.874  &   375.839  &    375.951  & 0.077  & 0.112  \\
 $G_{hadron},~\MeV$         &  1741.039  &  1741.268  &   1741.442  & 0.403  & 0.174 \\
 $\Gamma_Z,~\MeV$             &  2494.387  &  2494.636  &   2494.814  & 0.427  & 0.178 \\
 \hline
\end{tabular}
\caption{
Partial $G_{ij}$ and total $\Gamma_Z$ decay widths of the Z-boson
for sets of flags (IHVP,~IAMT4): (1,8) in comparison with (5,6) and (1,8) in comparison with (5,8).} 
\label{widths}
\end{table}
In
Table 
\ref{widths}  
we present the relevant numbers obtained with the latest update
of {\tt DIZET}  options
and previous actual options of these flags, i.e. 
(IHVP,~IAMT4): (1,8) in comparison with (5,6) and (1,8) in comparison with (5,8).
The main improvement comes from accounting previously missing bosonic $O(\alpha^2)$ corrections to the $Z\to b\bar{b}$ decay~\cite{Dubovyk:2016aqv}.

\subsubsection{The effective weak mixing angle $\sin^2\vartheta_{eff}$}

In Table~\ref{sineff} we illustrate
various options for flag AMT4 in
{\tt DIZET} v.~6.45
to estimate
$\sin^2\vartheta_{eff}$  in different channels.

\begin{table}[ht]
\centering
\begin{tabular}{|l|ccc|c|c|}
\hline         
(IHVP,~IAMT4)        &      (1,8)  &     (5,6)    &      (5,8)   &  $|\delta_{(1,8)-(5,8)}|\cdot 10^{3}$& $|\delta_{(5,6)-(5,8)}|\cdot 10^{3}$  \\
\hline
$\sin^2\vartheta_{eff}$ channel           &           &            &            & &   \\
\hline 
 $\nu,\bar\nu$    &0.231280    & 0.231149  & 0.231118    & 0.162  & 0.031  \\
 $e^+,e^-$        &0.231661    & 0.231530  & 0.231500    & 0.161  & 0.030  \\
 $\mu^+,\mu^-$    &0.231661    & 0.231530  & 0.231500    & 0.161  & 0.030  \\
 $\tau^+,\tau^-$  &0.231661    & 0.231530  & 0.231500    & 0.161  & 0.030  \\
 $u,\bar u$       &0.231555    & 0.231424  & 0.231393    & 0.162  & 0.031  \\
 $d,\bar d$       &0.231428    & 0.231297  & 0.231266    & 0.162  & 0.031  \\
 $c,\bar c$       &0.231555    & 0.231424  & 0.231393    & 0.162  & 0.031  \\
 $s,\bar s$       &0.231428    & 0.231297  & 0.231266    & 0.162  & 0.031  \\
 $b,\bar b$       &0.232895    & 0.232970  & 0.232732    & 0.163  & 0.238  \\
  \hline
\end{tabular}
\caption{
The effective weak mixing angle $\sin^2\vartheta_{eff}$ for all channels 
calculated for
sets of flags (IHVP,~IAMT4):
 (1,8) in comparison with (5,6) and (1,8) in comparison with (5,8).}
 \label{sineff}
\end{table}
The results for the channels  are similar. The largest shift comes from accounting previously missing bosonic $O(\alpha^2)$ corrections to $\sin^2\theta^b_{eff}$~\cite{Dubovyk:2016aqv}.
\clearpage

\subsubsection{Cross Sections} 

\begin{figure}[ht]
\centering
\label{fig_sigma}
\includegraphics[width=12cm]{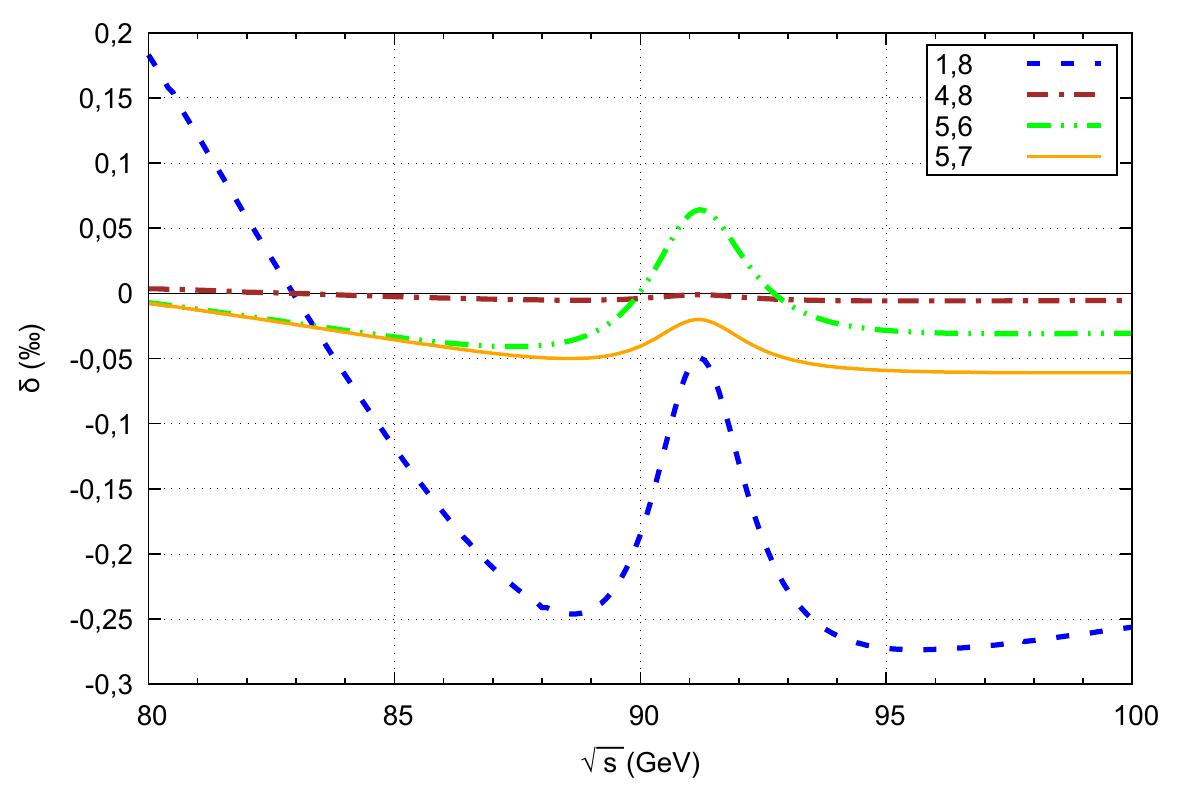}
\caption{Differences $\delta$ defined in Eq.~(\ref{eq:diffxsection}) for cross sections of $e^+e^-\to u \bar{u}$ for sets of flags (IHVP,~IAMT4): {(1,8), (4,8), (5,6), (5,7)} relative to the current best set {(5,8)}.}
\label{xs-uu}
\end{figure}
Using  the example of the channel $e^+e^-\to u \bar{u}$,   
Figure~\ref{xs-uu} shows  the cross section differences for  sets of  the flags (IHVP, IAMT) relative to the current best set (IHVP, IAMT) =(5,8);
 
\begin{equation} 
\delta = \displaystyle{\frac{\sigma (\rm{IHVP, IAMT}) - \sigma (5,8)}{\sigma (5,8)}}\cdot1000[\permil] .\nonumber \label{eq:diffxsection}
\end{equation}

The main influence on the result is the use of a  modern parametrization for the 
hadronic vacuum polarization. The effect of bosonic corrections is weaker. The results close to the $Z$ peak ($\pm$ 1 GeV) show that the relative shift is below $5\cdot10^{-5}$. 

\subsubsection{
Left-Right and Forward-Backward Asymmetries}
\begin{figure}[ht]
\centering
\includegraphics[width=12cm]{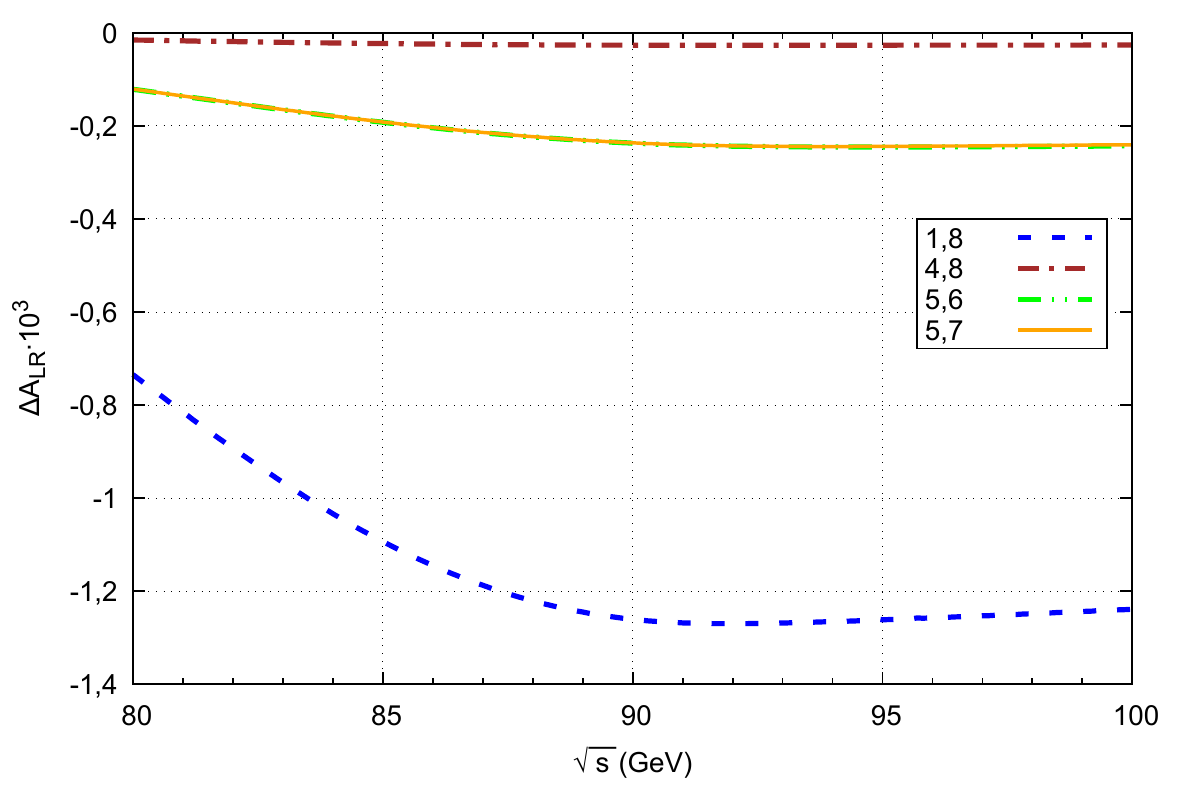}
\caption{Differences for left-right asymmetries of $e^+e^-\to u \bar{u}$ for sets of flags (IHVP,~IAMT4): {(1,8), (4,8), (5,6), (5,7)} from the current best set {(5,8)}.}
\label{fig_ALR}
\end{figure}

\begin{figure}[ht]
\centering
\includegraphics[width=12cm]{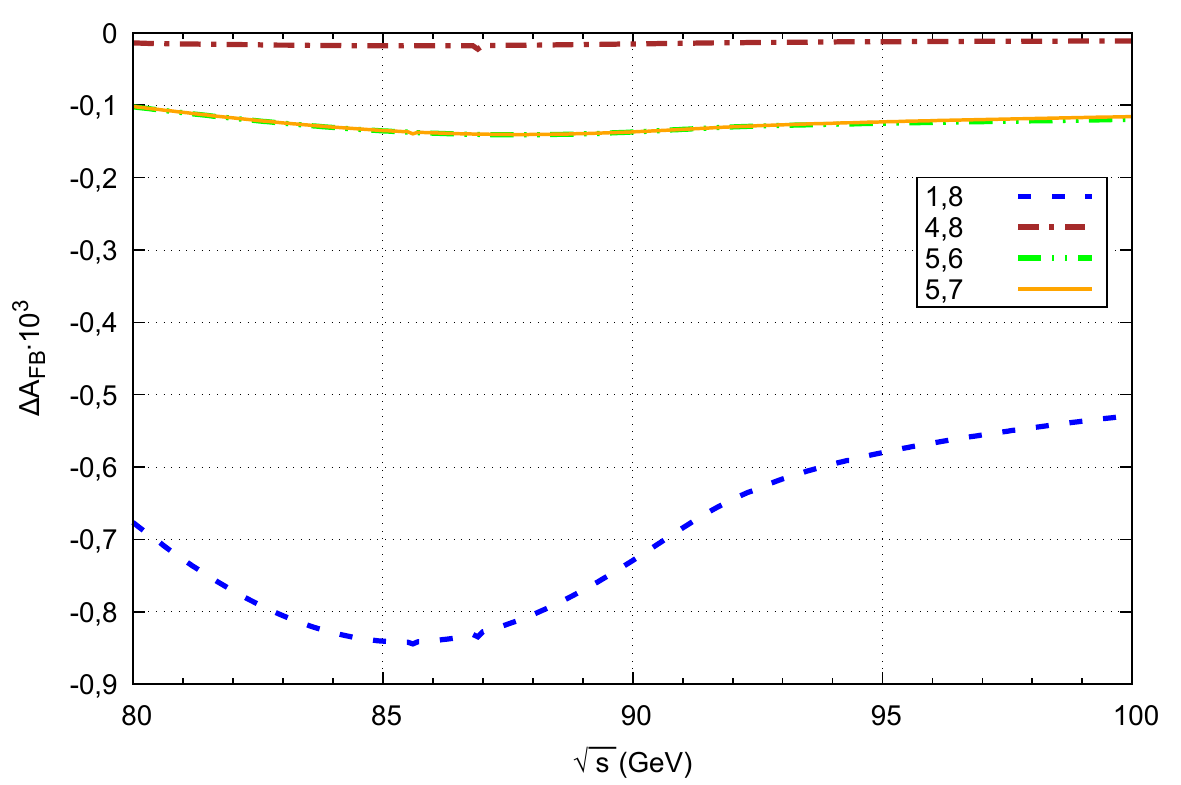}
\caption{Differences for forward-backward asymmetries of $e^+e^-\to u \bar{u}$ for sets of flags (IHVP,~IAMT4): {(1,8), (4,8), (5,6), (5,7)} from the current best set {(5,8)}.}
\label{fig_AFB}
\end{figure}

The  channel $e^+e^-\to u \bar{u}$ is  used to show in 
Figures~\ref{fig_ALR} and \ref{fig_AFB}  the  differences for the left-right asymmetries  and for the forward-backward asymmetries for sets of  the flags (IHVP, IAMT) compared with  the current best set (IHVP, IAMT) = (5,8);

\begin{equation} 
\Delta \rm{A} = \rm{A}(\rm{IHVP, IAMT})- \rm{A}(5,8). \nonumber
\end{equation}

As in the case of corrections to the cross section the main effect is due to changes in the  hadronic vacuum polarization treatment. Bosonic corrections of the $O(\alpha^2)$ order taken into account change the left-right asymmetry by no more than $3\cdot10^{-4}$ and the forward-backward asymmetry by no more than $2\cdot10^{-4}$.

\clearpage
\section{Benchmarks {\tt DIZET} v.~6.45}

Here, we provide a benchmark for the default set of parameters and flags. 
\begin{verbatim}
 DIZET flags, see routine DIZET for explanation:
  IHVP = 5 IAMT4 = 8
  Iqcd = 3 Imoms = 1
 Imass = 0 Iscre = 0
 Ialem = 3 Imask = 0
 Iscal = 0 Ibarb = 2
 IFtjr = 1 Ifacr = 0
 IFact = 0 Ihigs = 0
 Iafmt = 3 Iewlc = 1
 Iczak = 1 Ihig2 = 0
 Iale2 = 3 Igfer = 2
 Iddzz = 1 Iamw2 = 0
 Isfsr = 1 Idmww = 0
 Idsww = 0
    IBAIKOV =          2012
    IBAIKOV =          2012

 DIZET input parameters:
 ZMASS   91.187600000000003      TMASS    172.75999999999999     
 HMASS   125.25000000000000      WMASS    0.0000000000000000     
 DAL5H   0.0000000000000000      ALQED5   137.03599908400000     
 ALFAS   0.11790000000000000     

 DIZET results:
 SIN2TW     0.22340388419691781     
 WMASSsin   80.358790700232220     
 WMASS      80.358790700232220     
 DAL5H      2.7576193213462830E-002
 ALQED5     128.95030472145015     
 ALST       0.10755034917841029     
 ALPAS      0.11790000000000000     

 CHANNEL         WIDTH         RHO_F_R        RHO_F_T        SIN2_EFF
 -------        -------       --------       --------       --------
 nu,nubar       167.202       1.007963       1.007963       0.231118
 e+,e-           83.985       1.005219       1.005062       0.231500
 mu+,mu-         83.985       1.005219       1.005062       0.231500
 tau+,tau-       83.795       1.005219       1.005062       0.231500
 u,ubar         299.918       1.005812       1.005757       0.231393
 d,dbar         382.861       1.006733       1.006723       0.231266
 c,cbar         299.852       1.005812       1.005757       0.231393
 s,sbar         382.861       1.006733       1.006723       0.231266
 t,tbar           0.000       0.000000       0.000000       0.000000
 b,bbar         375.839       0.994198       0.994198       0.232732
 hadron        1741.442
 total         2494.814

 W-widths
 lept,nubar     678.981
 down,ubar     1410.986
 total         2089.967

 FF:
 RHO             (0.99876990486335659,-4.73600008701756652E-003)
 RHO      (0.99876990  -.00473600)
 KAPPA.I  (1.03606482  0.01353154)
 KAPPA.J  (1.04380408  0.01353154)
 KAPPA.IJ (1.08147655  0.02706307)
 AL_I(s)                 (129.37048577139927,1.9810219926255057)
 AL_5_I(s)               (129.36196765366330,1.9807610678870036)
 ****************************************************
\end{verbatim}

\section{Conclusions}

The new version 6.45 of the {\tt DIZET} electroweak library is described.
In this work, we benchmark the novel implementation
of two-loop $\alpha^2$ bosonic and fermionic radiative corrections
and several fits of the hadronic vacuum polarization 
$\Delta \alpha^{(5)}_{had}(M_{\sss Z})$ using public versions of the {\tt AlphaQED} code.
The presented numerical results show the impact of the new options
for cross sections,
left-right and forward-backward asymmetries.
We can see that the updates
are relevant for high-precision experiments at future electron-positron colliders. 
Compatibility with previous versions of the code is supported.
Predictions for observables and pseudobservables have been produced with the {\tt ZFITTER} program \cite{Bardin:1999yd,Arbuzov:2005ma}.
The code is available directly with the link:
\href{http://sanc.jinr.ru/users/zfitter/DIZET_v6.45.tgz} {DIZET6.45}. 
Recently the new electroweak library {\tt GRIFFIN} was created \cite{Chen:2022dow} in which comparisons with {\tt DIZET} are given. The new version of the discussed here {\tt DIZET} program can serve for further tuned comparisons in future high-precision studies.  
\\

{\bf{Note:}} The work done in this paper is motivated by the  HL-LHC studies and the LHC electroweak Working Group activity and is based on the implementation of the new version of the DIZET code released on December 2019 \cite{dizetv645} and public presentations of results  during EWG meetings in 2019 \cite{dizetws,dizetws2} and 2020 \cite{dizetws3}.

\section*{Acknowledgments}
This work has been supported in part by the Polish National Science Center (NCN) under grant 2017/25/B/ST2/01987. A.A., L.K. and V.Ye. are grateful for the support to RFBR grant N~20-02-00441.

\bibliographystyle{elsarticle-num-ID}
\bibliography{dizet645,2loops}

\end{document}